\documentclass[12pt]{iopart}
\usepackage{epsfig}
\begin{document}

\newcommand{\zb}{\mbox{$\rm Z^0$}}
\newcommand{\ccbar}{\mbox{$\rm c\overline{c}$}}
\newcommand{\bbbar}{\mbox{$\rm b\overline{b}$}}
\newcommand{\qqbar}{\mbox{$\rm q\overline{q}$}}
\newcommand{\bqbar}{\mbox{$\rm\overline{b}$}}
\newcommand{\bplus}{\mbox{$\rm B^+$}}
\newcommand{\bminus}{\mbox{$\rm B^-$}}
\newcommand{\bzero}{\mbox{$\rm B^0_d$}}
\newcommand{\bzerobar}{\mbox{$\rm\bar{B}^0_d$}}
\newcommand{\bes}{\mbox{$\rm B^0_s$}}
\newcommand{\bsbar}{\mbox{$\rm \bar{B}^0_s$}}
\newcommand{\bbary}{\mbox{$\rm \Lambda_b$}}
\newcommand{\bbarybar}{\mbox{$\rm \bar{\Lambda}_b$}}
\newcommand{\lambb}{\mbox{$\rm\Lambda_b^0$}}
\newcommand{\dstar}{\mbox{$\rm D^{*+}$}}
\newcommand{\ddstar}{\mbox{$\rm D^{**}$}}
\newcommand{\dzero}{\mbox{$\rm D^0$}}%
\newcommand{\bztodslv}{$\bzerobar\rightarrow\dstar\ell^-\bar{\nu}$}
\newcommand{\brbtodslv}{\mbox{$\rm Br(\bzerobar\rightarrow D^{*+}\ell^-\bar{\nu})$}}
\newcommand{\btodshlv}{$\rm \bar{B}\rightarrow\dstar h\,\ell^-\bar{\nu}$}
\newcommand{\btodsslv}{$\rm \bar{B}\rightarrow\ddstar\ell^-\bar{\nu}$}
\newcommand{\fbd}{\mbox{$f_{\rm B^0}$}}
\newcommand{\btoxulv}{\mbox{$\rm b\rightarrow X_u\ell\bar\nu$}}
\newcommand{\btoclv}{\mbox{$\rm b\rightarrow c\ell\bar\nu$}}
\newcommand{\vcb}{\mbox{$V_{\rm cb}$}}
\newcommand{\vub}{\mbox{$V_{\rm ub}$}}
\newcommand{\mvcb}{\mbox{$|V_{\rm cb}|$}}
\newcommand{\mvub}{\mbox{$|V_{\rm ub}|$}}
\newcommand{\bvcb}{\mbox{$V_{\rm cb}$}}
\newcommand{\fvcb}{\mbox{${\cal F}(1)|V_{\rm cb}|$}}
\newcommand{\bfvcb}{\mbox{${\cal F}(1)|V_{\rm cb}|$}}
\newcommand{\rhsq}{\mbox{$\rho^2$}}
\newcommand{\brhsq}{\mbox{$\rho^2$}}
\newcommand{\fw}{\mbox{${\cal F}(\omega)$}}
\newcommand{\kw}{\mbox{${\cal K}(\omega)$}}
\newcommand{\fone}{\mbox{${\cal F}(1)$}}
\newcommand{\mean}[1]{\langle{#1}\rangle}
\newcommand{\meanxe}{\mbox{$\mean{x_E}$}}
\newcommand{\bratio}[2]{\mbox{${\rm Br}(#1\rightarrow #2)$}}
\newcommand{\delm}{\mbox{$\Delta m$}}
\newcommand{\dms}{\mbox{$\Delta m_{\rm s}$}}
\newcommand{\etake}{\mbox{$\eta_{\rm ke}$}}
\newcommand{\wt}{\mbox{$\omega'$}}
\newcommand{\tauh}{\mbox{$\hat{\tau_1}$}}
\newcommand{\zetah}{\mbox{$\hat{\zeta_1}$}}
\newcommand{\taubz}{\mbox{$\rm\tau_{B^0}$}}
\newcommand{\taubp}{\mbox{$\rm\tau_{B^+}$}}
\newcommand{\taubs}{\mbox{$\rm\tau_{B_s}$}}
\newcommand{\delgams}{\mbox{$\Delta\Gamma_{\rm s}$}}
\newcommand{\bshort}{\mbox{$\rm B_s^{short}$}}
\newcommand{\blong}{\mbox{$\rm B_s^{long}$}}
\newcommand{\gshort}{\mbox{$\rm\Gamma_s^{short}$}}
\newcommand{\glong}{\mbox{$\rm\Gamma_s^{long}$}}
\newcommand{\tshort}{\mbox{$\rm\tau_{B_s}^{short}$}}
\newcommand{\tlong}{\mbox{$\rm\tau_{B_s}^{long}$}}
\def\bb{\rm b\bar{b}}
\def\bd{\rm B^0_d} 
\def\bq{\rm B^0_q}
\def\bs{\rm B^0_s}
\def\bdb{\rm \overline{B^0_d}} 
\def\bqb{\rm \overline{B^0_q}} 
\def\bmix{\rm B^0 \mbox{--} \overline{B^0}}
\def\bdmix{\rm B_d^0 \mbox{--} \overline{B_d^0}}
\def\bqmix{\rm B_q^0 \mbox{--} \overline{B_q^0}}
\def\bsmix{\rm B_s^0 \mbox{--} \overline{B_s^0}}
\def\dmd{\Delta m_d}
\def\dmq{\Delta m_q}
\def\dms{\Delta m_s}
\def\ips{\mbox{ps}^{-1}}
\def\vtd{V_{td}}
\def\vts{V_{ts}}
\def\Zbb{\rm Z^0 \rightarrow b\,{\overline b}}
\newcommand{\rPLB}[3] {{#2} {\it Phys.~Lett.} B {\bf #1} #3}
\newcommand{\rPRL}[3] {{#2} {\it Phys.~Rev.~Lett} {\bf #1} #3}
\newcommand{\rPRD}[3] {{#2} {\it Phys.~Rev.} D {\bf #1} #3}
\newcommand{\rPRP}[3] {{#2} {\it Phys.~Rep.} {\bf #1} #3}
\newcommand{\rNIM}[3] {{#2} {\it Nucl.~Instrum. Methods} {\bf #1} #3}
\newcommand{\rNPB}[3] {{#2} {\it Nucl.~Phys.} B {\bf #1} #3}
\newcommand{\rCPC}[3] {{#2} {\it Comp.~Phys. Comm.} {\bf #1} #3}
\newcommand{\rZPC}[3] {{#2} {\it Z.~Phys.} C {\bf #1} #3}
\newcommand{\rJPH}[3] {{#2} {\it J.~Phys.} {\bf #1} #3}
\newcommand{\rEPJ}[3] {{#2} {\it Eur.~Phys.\ J.} C {\bf #1} #3}
\newcommand{\rIJA}[3] {{#2} {\it Int.~J.~Mod.~Phys.} A {\bf #1} #3}
\newcommand{\rSJN}[3] {{#2} {\it Sov.~J.~Nucl.~Phys.} {\bf #1} #3}
\newcommand{\rhepostfig}[5]{
\begin{figure}[tbp]
\setlength{\epsfxsize}{#4\hsize}
\hspace*{#3\hsize} \epsfbox{#1}
\caption{\label{#2}#5}
\end{figure}
}
%
%
%
\title{B physics at the ${\rm Z^0}$ resonance}

\author{Richard Hawkings\dag and St\'ephane Willocq\ddag  
\footnote[3]{E-mail: {\tt richard.hawkings@cern.ch}
  and {\tt willocq@physics.umass.edu}}
}

\address{\dag\ CERN-EP division, CH 1211 Geneva 23, Switzerland}

\address{\ddag\ Physics Department, University of Massachusetts,
 Amherst, MA 01003, USA}

\begin{abstract}
B physics results from $e^+ e^-$ annihilation at the ${\rm Z^0}$
resonance are reviewed.
A vast program is summarised, including the study of
\bplus, \bzero, \bes\ and b baryon lifetimes,
the time dependence of $\bdmix$ and $\bsmix$ oscillations,
the width difference in the $\bsmix$ system,
and the measurements of the magnitudes of the CKM
matrix elements \vcb\ and \vub.
\end{abstract}




\section{Introduction}

The production of b hadrons in $e^+e^-$ annihilation at the \zb\ resonance
offers an excellent opportunity to study their properties in detail. The 
four LEP experiments have each accumulated 0.9 million $\zb\rightarrow\bbbar$
events, and the SLD experiment at the SLAC linear collider has collected
a further 120\,000 \bbbar\ events with a polarised $e^-$ beam and excellent 
secondary vertex resolution from a pixel-based vertex detector.
These data samples have been used to perform a huge program of b physics, 
including the studies of b hadron lifetimes, \bzero\ and \bes\ mixing, the
$\bsmix$ width difference and measurements of the CKM matrix element
magnitudes \mvcb\ and \mvub\ described below.

\section{b hadron lifetimes}

The lifetimes of b hadrons depend both on the strength of the coupling
of the b quark to the lighter c and u quarks, and on the dynamics of the
b hadron decay. The spectator model prediction that all b hadron lifetimes
are equal is modified by effects dependent on the flavour
of the light quark(s) in the hadron. These effects can be calculated using
Heavy Quark Expansion tools, and the precise measurement
of b hadron lifetimes provides an important test of this theory.

Experimentally, the measurement of b hadron lifetimes requires the isolation
of a sample of the particle under study (both from other b hadrons and non
b background), the reconstruction of the proper decay time of each b hadron,
and the detailed understanding of backgrounds and systematic biases.
Different approaches have been used, and will be discussed in more detail 
below.

\subsection{\bplus\ and \bzero\ lifetimes}\label{rhs:taubp}

The purest samples of \bplus\ and \bzero\ hadrons are obtained via 
exclusive reconstruction, for example semileptonic
decays of the form $\bminus\rightarrow\dzero\ell^-\bar\nu$ and
\bztodslv\ followed by $\dstar\rightarrow\dzero\pi^+$. The 
\dzero\ can be reconstructed in various exclusive decay modes, for
example $\dzero\rightarrow\rm K^-\pi^+$, $\rm K^-\pi^+\pi^0$,
$\rm K^-\pi^+\pi^-\pi^+$ and $\rm K^0_s\pi^+\pi^-$. A recent
ALEPH analysis \cite{alephexcl} obtains 1880 \bzero\ and 2856 \bplus\
candidates at 85\,\% and 80\,\% purity from their recently reprocessed
LEP1 data sample. The sample purities are limited both by combinatorial
background (random combinations of tracks faking a \dzero\ or \dstar), and
by cross-contamination of the \bplus\ and \bzero\ samples. The latter
is caused mainly by decays of the form 
$\bminus\rightarrow\ddstar^0\ell^-\bar\nu$ and 
$\bzerobar\rightarrow\ddstar^+\ell^-\bar\nu$, where the orbitally excited
\ddstar\ decays to $\dstar\pi$ or $\dzero\pi$. Fully exclusive reconstruction
of particular hadronic decay modes, for example 
$\bminus\rightarrow\dzero\pi^-$ and $\bzerobar\rightarrow\dstar\pi^-$
has also been used by ALEPH \cite{alephexclh}, but the reconstructed
sample sizes are too small to produce competitive lifetime measurements.

Much larger samples of b hadrons can be obtained using vertex-based b tagging,
exploiting the silicon microvertex detectors installed in the LEP and 
SLD experiments. Events with a clear separated secondary vertex can be
classified as charged (\bplus) or neutral (\bzero, \bes\ or \bbary)
by reconstructing the charge of the secondary vertex from the associated
tracks. This technique results in large high purity \bplus\ samples, but
the \bzero\ sample is limited by irreducible contamination from 
\bes\ and \bbary. The selection of clear secondary vertices results in 
a bias towards large b hadron lifetimes, which must be controlled.
In the OPAL analysis \cite{opaltop}, this is handled using an `excess decay
length' technique, finding the minimum decay length for each event that would
still result in it being selected, whilst the higher resolution silicon 
detectors of DELPHI and especially SLD allow them simply to cut events
with small proper times and fit an unbiased exponential beyond the cut
\cite{delsldtop}.
In all these analyses, the charged/neutral separation is enhanced using
additional information, for example by reconstructing the flavour (b or 
\bqbar) of the b hadron in the opposite event hemisphere, or by using 
identified kaons or protons to suppress \bes\ and \bbary\ background.
These techniques result in samples of tens of thousands of charged
and neutral b decays, even for SLD where the high performance vertex detector
more than compensates for the lower initial number of \zb\ compared to LEP.

A third technique to measure the \bzero\ lifetime involves exploiting the
small energy release in the decay $\dstar\rightarrow\dzero\pi^+$, which 
means that in \bztodslv\ decays the $\pi^+$ is produced with very small
transverse momentum with respect to the \dzero\ direction. The
\dzero\ direction can be reconstructed inclusively by weighting 
tracks and clusters according to kinematic and vertexing information, 
without the need to explicitly reconstruct the decay modes of the \dzero.
\bztodslv\ events are selected by looking for charged pions with small 
transverse momentum and sign opposite to that of the lepton, and the 
combinatorial background is controlled using events with same sign pions and 
leptons. Additional background contamination comes from \btodsslv\ decays
in the same way as for exclusively reconstructed \dzero. Using this 
technique, DELPHI, L3 and OPAL \cite{dell3tbz,opaltbz}
have reconstructed samples of
around 5000 \bzero\ decays and measured the lifetime.

\begin{figure}[tb]
\begin{centering}
  \epsfxsize8cm
  \epsfbox{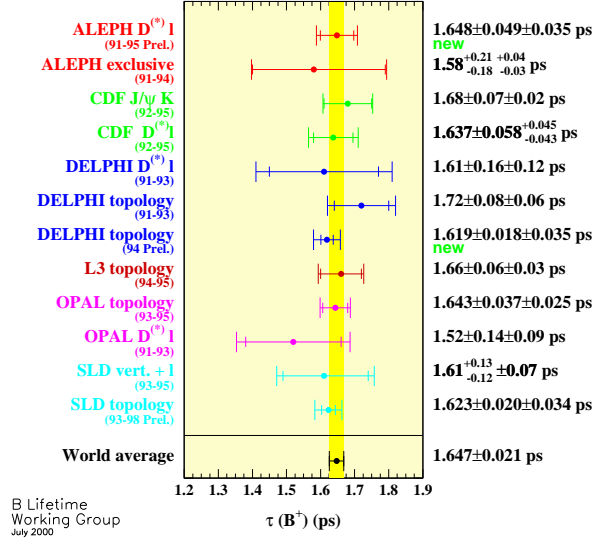}
  \caption{\label{rhf:bplife}
Summary of the \bplus\ lifetime measurements from
LEP, CDF and SLD.}
\end{centering}
\end{figure}

\begin{figure}[tb]
\begin{centering}
  \epsfxsize8cm
  \epsfbox{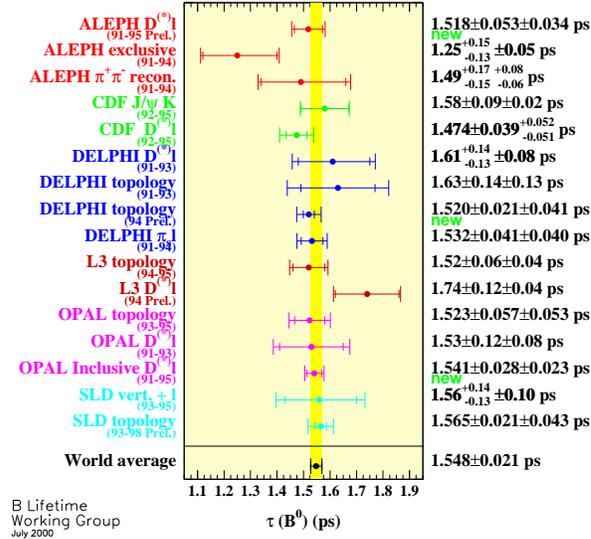}
  \caption{\label{rhf:bzlife}
Summary of the \bzero\ lifetime measurements from
LEP, CDF and SLD.}
\end{centering}
\end{figure}

A summary of all \bplus\ and \bzero\ lifetime measurements is given in 
Figures~\ref{rhf:bplife} and \ref{rhf:bzlife} \cite{lephfst}.
The \bplus\ measurements are
dominated by the topological secondary vertex analyses from DELPHI, OPAL and
SLD \cite{opaltop,delsldtop} 
whilst the most accurate \bzero\ measurements come from the 
OPAL inclusive \bztodslv\ analysis \cite{opaltbz} and the 
DELPHI and SLD topological measurements \cite{delsldtop}.


\subsection{\bes\ and \bbary\ lifetimes}\label{rhs:taubs}

\begin{figure}[tbp]
\parbox{0.48\textwidth}
{\setlength{\epsfxsize}{1.1\hsize}\epsfbox{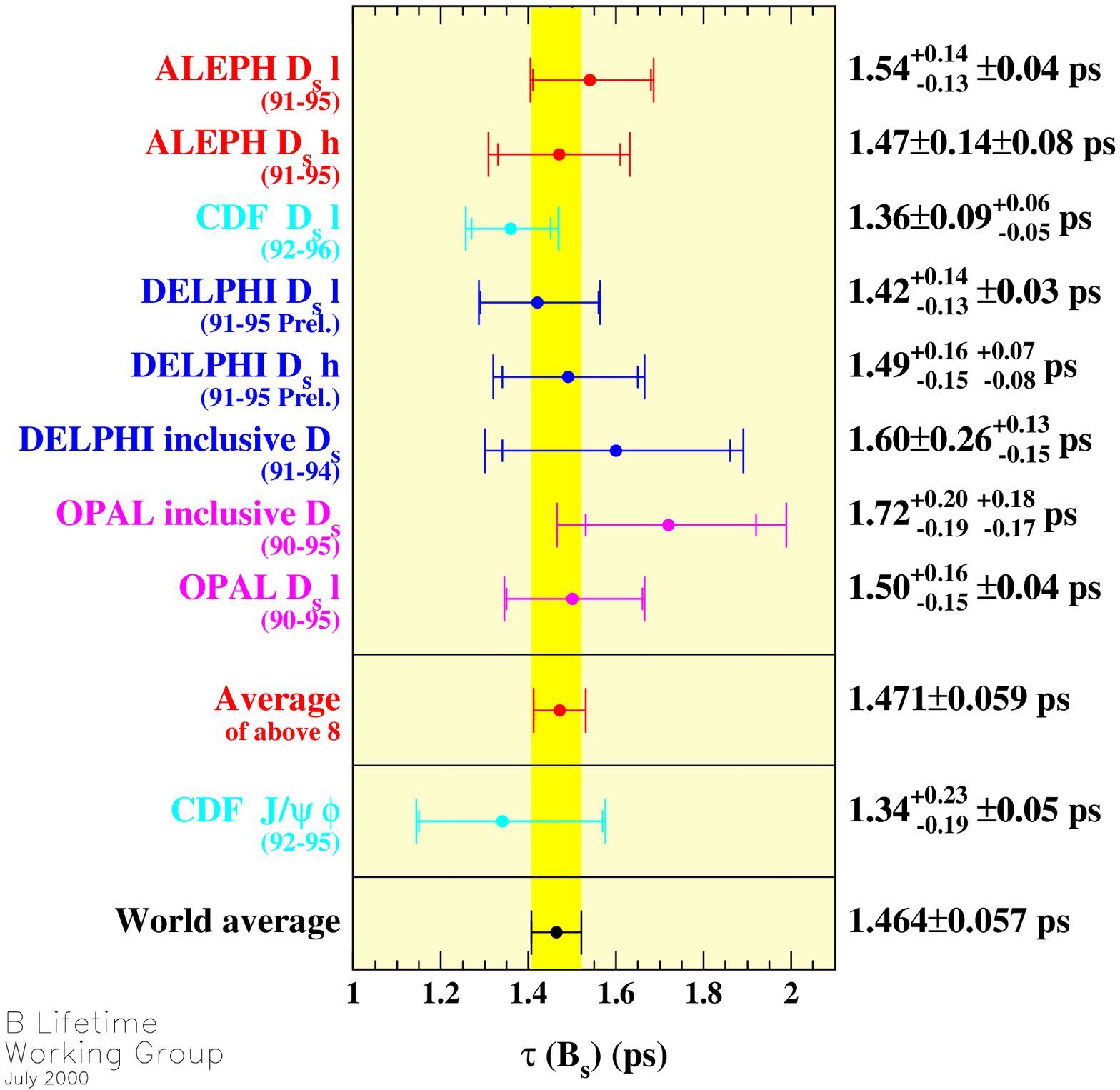}}
\parbox{0.48\textwidth}
{\setlength{\epsfxsize}{1.1\hsize}\epsfbox{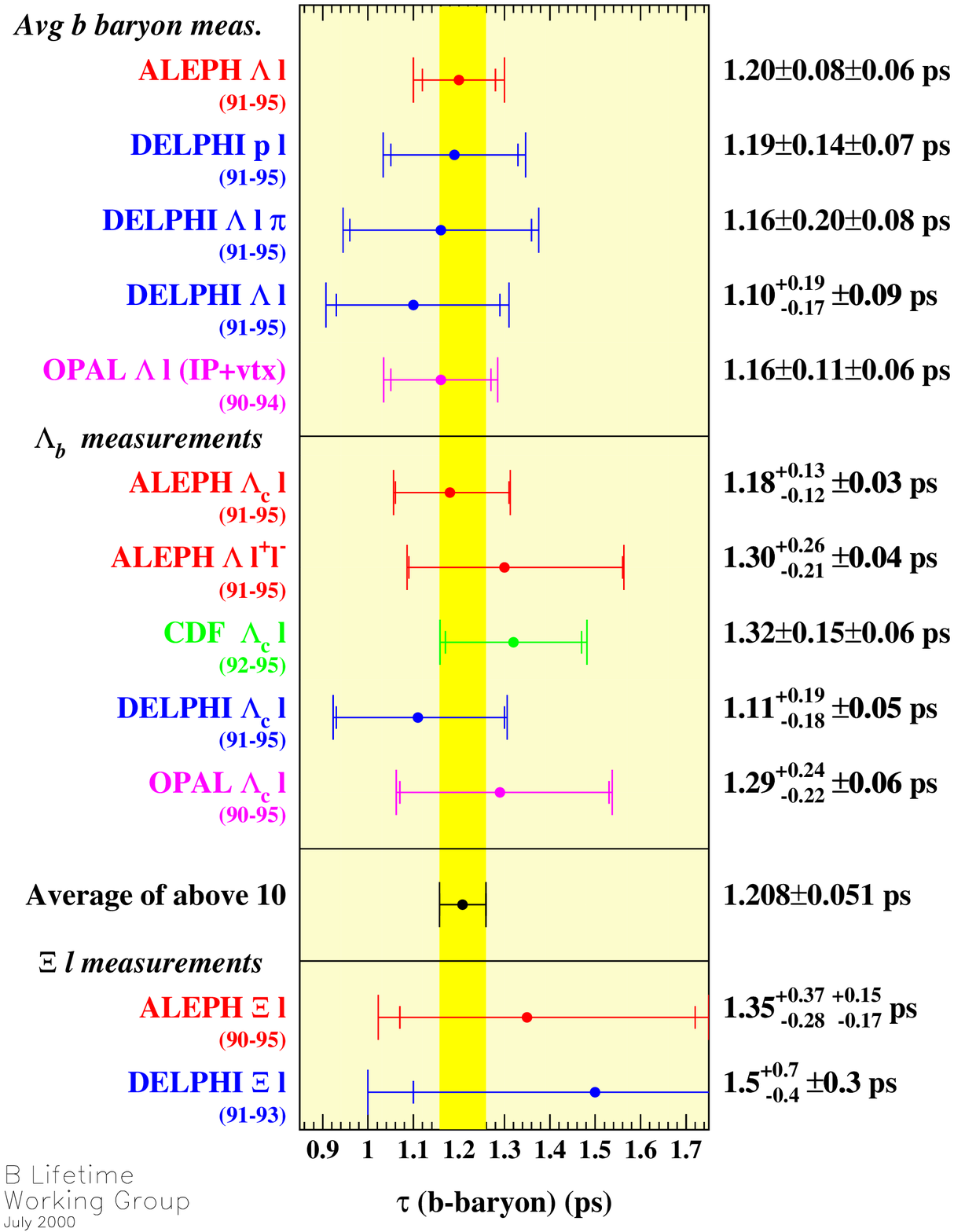}}
\caption{\label{rhf:bsbblife}
Summary of the \bes\ and b baryon lifetime measurements from
LEP and CDF.}
\end{figure}

In $\zb\rightarrow\bbbar$ decays, only about 10\,\% of b quarks hadronise
to form \bes\ mesons, and a further 10\,\% form b baryons \cite{lephfst}.
The smaller numbers of hadrons compared to \bplus\ and \bzero\ mean that 
only exclusive reconstruction techniques
can be used to isolate pure samples, and these samples are rather 
small---typically less than a thousand events. The available measurements
are summarised in Figure~\ref{rhf:bsbblife} \cite{lephfst}. The most precise 
\bes\ lifetime results come from semileptonic 
$\bes\rightarrow\rm D_s\ell\bar\nu$,
augmented by more inclusive $\bes\rightarrow\rm D_sX$ hadronic decay analyses
with lower purity. The b baryon decay results are split into two 
groups---those from inclusive $\Lambda\ell$ correlations, and those in
which an intermediate charm baryon $\Lambda_{\rm c}^+$ is 
exclusively reconstructed.
The latter is expected to be enhanced in the lightest b-flavoured baryon
$\rm\Lambda_b^0$. There are also measurements of the $\rm\Xi_b$ lifetime
via $\Xi\ell$ correlations, but these are of very limited statistical 
precision \cite{xiblife}.

\subsection{Lifetimes summary}

The lifetime results can be most usefully compared with theory by studying
the lifetime ratios $\taubp/\taubz$, $\taubs/\taubz$, {\em etc\/}. These
are shown in Figure~\ref{rhf:avsum}, together with the corresponding 
theoretical predictions \cite{lephfst}. Whilst the meson lifetime ratios
are in excellent agreement, the b baryon lifetime is significantly smaller
than expected. The experimental results have been stable for several years,
and it remains to be seen whether they can be reconciled with the existing
theoretical picture, or are pointing to a more fundamental problem.
The LEP and SLD lifetime analyses are now almost complete, so further
significant experimental improvement can only be expected from the
$\rm\Upsilon(4S)$ B factories (for \bplus\ and \bzero\ only)
and Run II of the Tevatron.

\rhepostfig{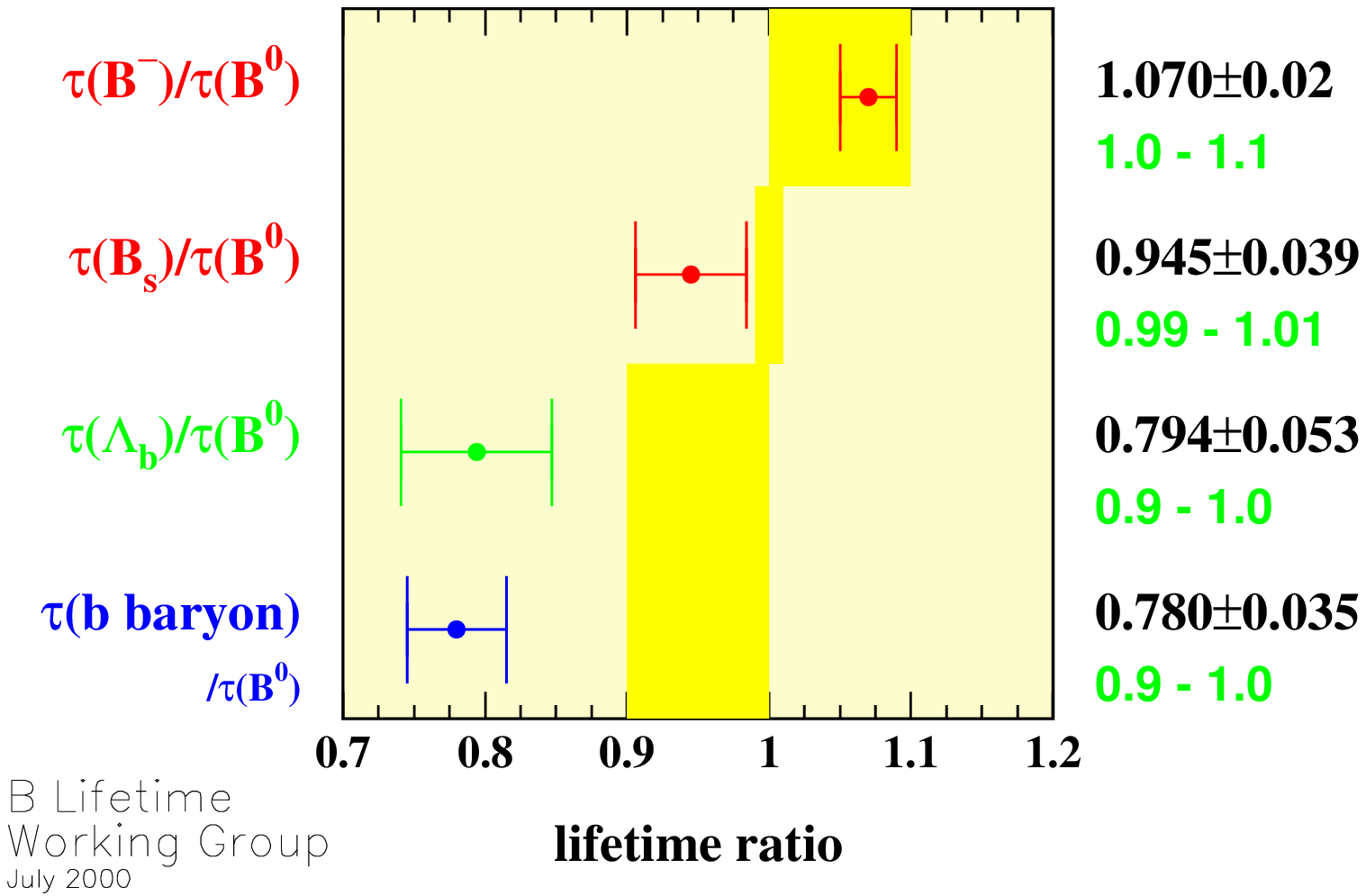}{rhf:avsum}{0.2}{0.7}
{Summary of experimental measurements of lifetime ratios, together with
theoretical expectations.}

\section{B mixing}

\subsection{Introduction}

  The $\bqmix$ system consists of $\bq$ and $\bqb$ flavour
eigenstates, which are superpositions of heavy and light mass
eigenstates ($q= d$ and $s$ for $\bd$ and $\bs$ mesons, respectively).
The difference in mass $\dmq$ between these eigenstates
leads to ${\rm B_q^0 \leftrightarrow \overline{B_q^0}}$ oscillations
with a frequency (see Ref.~\cite{Buras})
\begin{equation}
  \dmq = \frac{G_F^2}{6\pi^2} m_{B_q} m_t^2 F(m_t^2 / m_W^2) f_{B_q}^2 B_{B_q}
         \eta_{QCD} \left| V_{tb}^\ast V_{tq} \right|^2,
  \label{eq_dmq}
\end{equation}
where $G_F$ is the Fermi constant, $m_{B_q}$ is the $\bq$ hadron mass,
$m_t$ is the top quark mass, $m_W$ is the $W$ boson mass,
$F$ is a function defined in Ref.~\cite{Inami},
and $\eta_{QCD}$ is a perturbative QCD parameter.
The parameter $B_{B_q}$ and the decay constant $f_{B_q}$ parameterize hadronic
matrix elements.
Much of the interest in B mixing stems from the fact that
a measurement of the $\bd$ ($\bs$) oscillation frequency
allows the magnitude of the
CKM matrix element $V_{td}$ ($V_{ts}$) to be determined,
see equation~(\ref{eq_dmq}).
However, this is limited by an uncertainty of about 20\% in the product
$f_{B_q} \sqrt{B_{B_q}}$~\cite{Bernard}.
Uncertainties are reduced for the ratio
\begin{equation*}
  \frac{\dms}{\dmd} = \frac{m_{B_s} f_{B_s}^2 B_{B_s}}{m_{B_d} f_{B_d}^2 B_{B_d}}
                      \left|\frac{V_{ts}}{V_{td}}\right|^2
                    = \frac{m_{B_s}}{m_{B_d}}~(1.16 \pm 0.05)^2 \left|\frac{V_{ts}}{V_{td}}\right|^2~,
\end{equation*}
which indicates that the ratio $|V_{ts}/V_{td}|$ can be determined
with an uncertainty as small as 5\%~\cite{Bernard}.
In the Wolfenstein parameterization of the CKM matrix, we have
$\dmd \propto |\vtd|^2 \simeq A^2 \lambda^6 [(1 - \rho)^2 + \eta^2]$
and $\dms \propto |\vts|^2 \simeq A^2 \lambda^4$,
where $\lambda = 0.2224 \pm 0.0020$ and $A = 0.83 \pm 0.03$~\cite{Faccioli},
but $\rho$ and $\eta$ are not well known.
As a result, studies of $\bd$ and $\bs$ mixing provide one of the
strongest constraints on the CKM unitarity triangle parameters $\rho$
and $\eta$, thus constraining CP violation in the Standard Model.

  Experimental studies require two main ingredients:
(i) reconstruction of the $\bq$ decay and its proper time,
(ii) determination of the $\bq$ or $\bqb$ flavour at both production and
decay to classify the decay as either `mixed' (if the tags disagree)
or `unmixed' (otherwise).
The significance for a $\bq$ oscillation signal can be approximated
by~\cite{Moser}
\begin{equation}
  S = \sqrt{\frac{N}{2}}\: f(B^0_q)\: \left[1 - 2\, w\right]\:
      e^{-\frac{1}{2} (\dmq \sigma_t)^2} ,
  \label{eq_signif}
\end{equation}
where $N$ is the total number of decays selected,
$f(\bq)$ is the fraction of $\bq$ mesons in the selected sample,
$w$ is the probability to incorrectly tag a decay as mixed or unmixed
(i.e. the mistag rate)
and $\sigma_t$ is the proper time resolution.
The proper time resolution depends on both the decay length resolution
$\sigma_L$ and the momentum resolution $\sigma_p$ according to
$\sigma_t^2 = (\sigma_L / \gamma\beta c)^2 + (t\, \sigma_p/p)^2$.
Based on the Wolfenstein parameterization, we see that
$\dms / \dmd \simeq 1 / \lambda^2$, which is of order of 20
(the other Wolfenstein parameters are of order 1).
Therefore, $\bs$ oscillations are expected to be much more rapid
than $\bd$ oscillations.
The ability to resolve such rapid oscillations thus requires excellent
decay length and momentum resolution, and benefits from having
a low mistag rate and a high $\bs$ purity.

\subsection{$\bd$ mixing}

  Measurements of the $\bd$ oscillation frequency have been performed
by ALEPH, DELPHI, L3, OPAL and SLD (see Ref.~\cite{BOSCWG}).
Most of these
rely on semileptonic B decays to provide a sample with high $\bd$ purity
and excellent decay flavour tag via the charge of the decay lepton:
a negatively (positively) charged lepton tags a $\bdb$ ($\bd$) decay.
One of the main challenges is to suppress leptons from cascade
$(b \to c \to l^+)$ decays which contribute ``wrong-sign'' leptons;
direct leptons come from $(b \to l^-)$ transitions.
Such suppression is usually achieved with a cut on $p_T$, the
momentum of the lepton transverse to the jet axis, or with a combination
of variables, e.g. using a neural network algorithm.

  Reconstruction methods fall in three general categories.
(i) nearly exclusive: a high purity sample is extracted
by selecting $\bzero \to {\rm D}^{(\ast) -} l^+ \nu_l$ decays
with subsequent 
${\rm D}^{\ast -} \to \overline{\rm D^0} \pi^-$
where $\overline{\rm D^0} \to K^+ \pi^- (\pi^0)$ or
$\overline{\rm D^0} \to K^+ \pi^- \pi^+ \pi^-$;
(ii) ${\rm D}^\ast$ inclusive: the previous decay chain can be
partially reconstructed with a much higher efficiency
by selecting the direct lepton and
the slow pion from the ${\rm D}^\ast$ decay, and by reconstructing
the ${\rm D^0}$ decay inclusively
(this is the same method described in Sec.~\ref{rhs:taubp});
(iii) fully inclusive: a lepton with high momentum $p$ and $p_T$
combined with an inclusive B and/or D vertex reconstruction.

\begin{figure}
\begin{center}
  \epsfxsize10cm
  \epsfbox{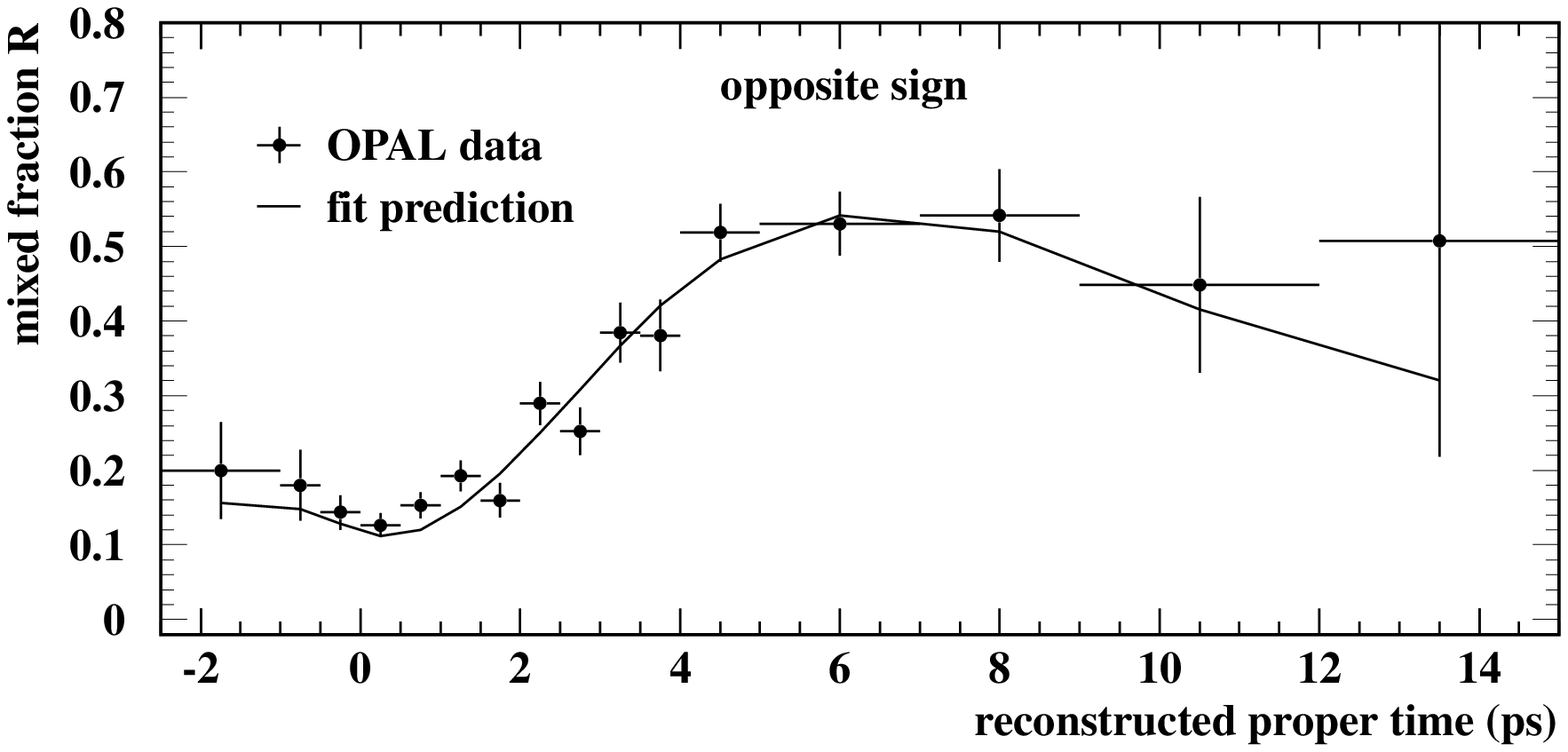}
\end{center}
\caption{Fraction of events tagged as mixed as a function of reconstructed
         proper time for the sample with opposite sign lepton and pion tracks.}
\label{fig_mixf_opal}
\end{figure}

The most precise single measurement performed at the \zb\ resonance
is the inclusive $D^\ast$ analysis by OPAL~\cite{opaltbz}:
high statistics samples of same-sign and opposite-sign lepton-pion pairs
are selected, with the same-sign pairs serving to constrain the
combinatorial background.
A clear oscillation signal is observed in the fraction of opposite-sign
events tagged as mixed (Fig.~\ref{fig_mixf_opal}) and a value of
$\dmd = 0.497 \pm 0.024 (stat) \pm 0.025 (syst)~\ips$ is extracted.

  All available measurements have been averaged to extract
a world average value of $\dmd = 0.487 \pm 0.014~\ips$~\cite{BOSCWG}.
This average is currently dominated
by LEP measurements but will soon be dominated by B Factory measurements,
thanks to much higher statistics.

\subsection{$\bs$ mixing}

  The study of the time dependence of $\bs$ mixing proceeds along the same
lines as that of $\bd$ mixing described earlier.
However, only about 10\% of b quarks fragment into $\bs$ mesons, as compared
with about 40\% into $\bd$ mesons. Furthermore, the $\bs$ oscillation frequency
is expected to be much larger than that for $\bd$ oscillations.
The search for $\bs$ oscillations thus presents a major experimental
challenge.
To address this, sophisticated analyses have been
developed with an emphasis on lowering the mistag rate, increasing
the $\bs$ purity and improving the proper time resolution, all of which
affect the sensitivity to $\bs$ mixing, see equation (\ref{eq_signif}).

  Tagging of the production flavour generally combines a number
of different methods.
The single most powerful tag exploits the large polarized forward-backward
asymmetry in $\Zbb$ decays. This tag is available at SLD thanks to the
large electron beam polarization ($P_e \simeq 73\%$).
A left- (right-) handed incident electron tags the quark
produced in the forward
hemisphere as a b ($\overline{\rm b}$) quark.
This method yields a mistag rate of $~28\%$ with nearly 100\% efficiency.
Tags used in all analyses rely on charge information from the event hemisphere
opposite that of the $\bs$ candidate:
(i) charge of lepton from the direct transition $b \to l^-$,
(ii) momentum-weighted jet charge,
(iii) secondary vertex charge,
(iv) charge of secondary vertex kaon from the dominant transition
$b \to c \to s$,
(v) charge dipole of secondary vertex (SLD only).
Other tags from the same hemisphere as the $\bs$ candidate are also used:
(i) unweighted (or weighted) jet charge, and (ii) charge of fragmentation kaon
accompanying the $\bs$.
These various tags are combined on an event-by-event basis to yield an
overall mistag rate of 20-25\%, depending on the particular analysis.

  The analyses differ in the way the $\bs$ decay is reconstructed
and thus in the
way the decay flavour is determined. Three general classes can be identified:
inclusive, semi-exclusive and fully exclusive. Inclusive analyses benefit
from the
large available statistics but suffer from low $\bs$ purity, whereas more
exclusive analyses benefit from high purity and resolution but suffer from
the lack of statistics
(this is particularly true for the fully exclusive analyses).
Several analyses are discussed below to highlight these differences.

  Inclusive analyses have been performed by ALEPH, DELPHI, OPAL and SLD.
The charge dipole analysis is an example of fully inclusive method introduced
by SLD~\cite{SLDbsmix}.
It aims to reconstruct the b hadron decay chain topology.
This method takes full advantage of the superb decay length resolution
of the SLD CCD pixel vertex detector to separate secondary tracks (from the B
decay point) from tertiary tracks (from the D decay point).
The decay length resolution is parametrized by the sum of two Gaussians with
$\sigma_L = 72~\mu$m (60\% fraction) and $265~\mu$m (40\%),
whereas the momentum resolution is parametrized with
$\sigma_p / p = 0.07$ (60\%) and 0.21 (40\%).
A ``charge dipole'' $\delta Q$ is defined as the distance between
secondary and tertiary
vertices signed by the charge difference between them such that
$\delta Q > 0$ ($\delta Q < 0$) tags \bzerobar (\bzero) decays.
The average decay flavour mistag rate is estimated to be 24\% and is
mostly due to decays producing two charmed hadrons.
A sample of 8556 decays is selected with a $\bs$ purity estimated to be 15\%
(higher than the production rate of 10\% due to the fact that
only neutral decays are selected).

  The most sensitive inclusive analysis was performed
by ALEPH~\cite{ALEPHbsmix1}
and aims to reconstruct semileptonic B decays.
In this analysis, the D decay vertex is reconstructed inclusively and
a resultant D track is vertexed with the lepton and the b hadron
direction (from the jet direction) to form a B decay vertex.
Fairly loose cuts are used at the various stages of the analysis to
obtain a high statistics sample of 74026 events.
The analysis relies on several neural network algorithms to perform
the following tasks:
production flavour tagging, $\bb$ event selection, direct lepton selection,
and $\bs$ fraction enhancement.
To maximize sensitivity to $\bs$ oscillations the analysis
incorporates all the information event by event since this helps
identify events with increased sensitivity.

  Semi-exclusive analyses have been performed by ALEPH, DELPHI, OPAL and SLD.
$\bs$ decays are partially reconstructed in the modes
$\bes \to {\rm D_s^-} l^+ \nu_l X$ and
$\bes \to {\rm D_s^- h^+} X$,
where h represents any charged hadron (or system of several hadrons)
and the ${\rm D_s^-}$ meson decay is either fully or partially reconstructed
in the modes ${\rm D_s^-} \to \phi\pi^-$, $K^{\ast 0} K^-$, $K^0 K^-$,
$\phi\pi^-\pi^+\pi^-$, $\phi l^- \overline{\nu_l}$, {\em etc}.

  The most sensitive semi-exclusive analysis performed by DELPHI selects
436 ${\rm D_s^-} l^+$ events~\cite{DELPHIbsmix1}.
The small statistics is compensated by the
high $\bes \to {\rm D_s^-} l^+\nu_l$ purity, estimated to be $\sim 53\%$,
and the good decay length and momentum resolution,
$\sigma_L = 200~\mu$m (82\% fraction) and $670~\mu$m (16\%),
$\sigma_p / p = 0.07$ (82\%) and 0.16 (16\%).
Analyses selecting $D_s^- h^+$ final states benefit from
higher statistics but are less sensitive than those selecting
$D_s^- l^+$ states because of lower $\bs$ purity and worse proper time
resolution.

  Finally, fully exclusive analyses have been performed
by ALEPH~\cite{ALEPHbsmix2}
and DELPHI~\cite{DELPHIbsmix2} via the modes
$\bes \to {\rm D_s^-}\pi^+$, ${\rm D_s^-} a_1^+$,
$\overline{\rm D^0} K^- \pi^+$,
and $\overline{\rm D^0} K^- a_1^+$, where the ${\rm D_s^-}$
and $\overline{\rm D^0}$
are fully reconstructed.
The number of decay candidates is 50 for ALEPH and 44 for DELPHI with signal
purities of approximately 40\% and 50\%, respectively.
The main advantage of this method is its excellent proper time resolution
with a negligible contribution from momentum resolution $\sim 0.5\%$.
As a result, $\sigma_t$ does not grow with increasing proper time $t$ and thus
the oscillation amplitude is not damped as $t$ increases.
Due to limited statistics, this method is not competitive with respect
to the inclusive and semi-exclusive methods described above.
However, this is the method of choice for future studies of $\bs$ oscillations
at hadron colliders.

\begin{figure}
\begin{center}
  \epsfxsize11cm
  \epsfbox{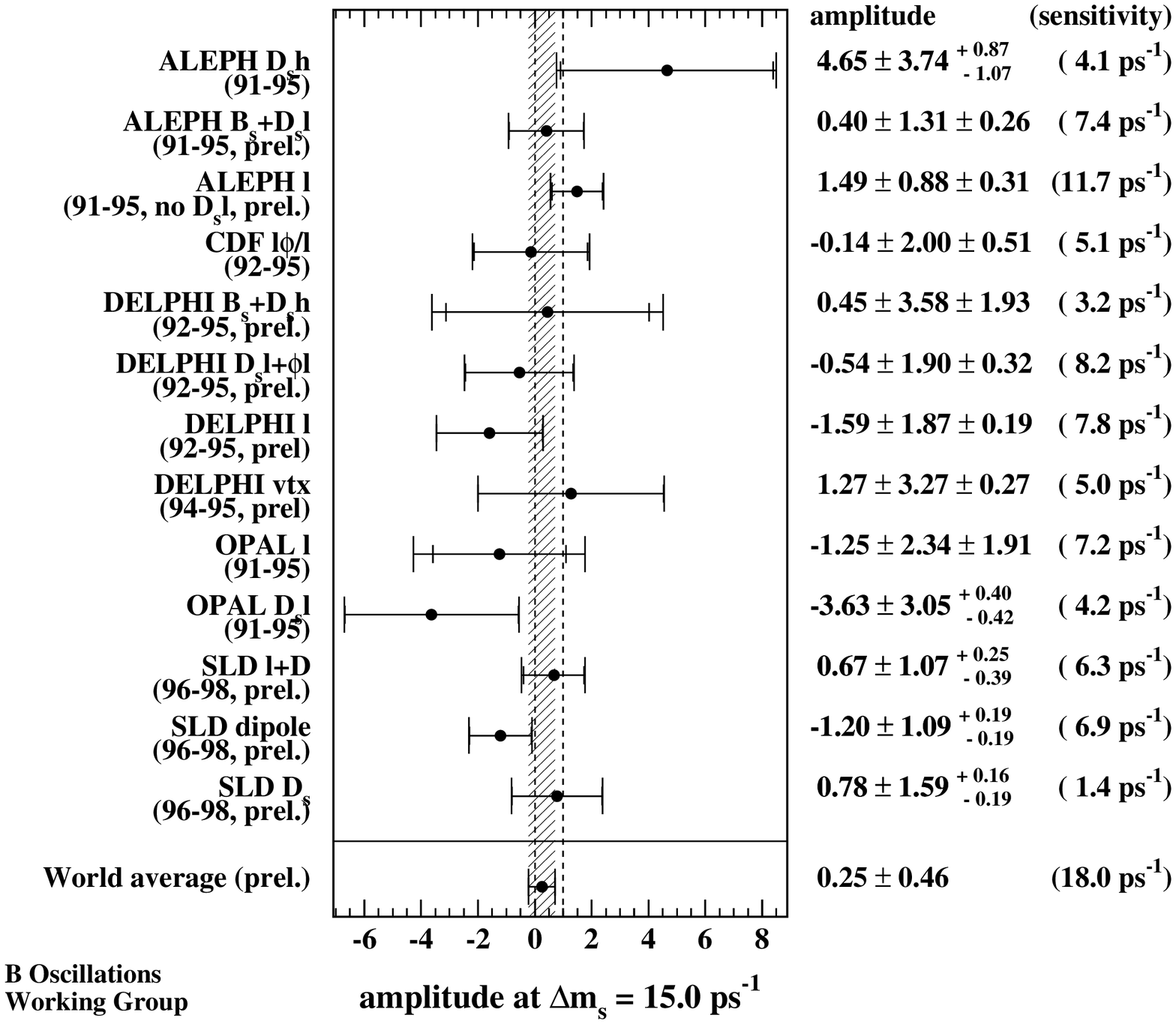}
\end{center}
\caption{Measurements of the $\bs$ oscillation amplitude
  for $\dms = 15~\ips$. }
\label{fig_ampW}
\end{figure}

  Studies of the time dependence of $\bs$ mixing are carried out
with the ``amplitude method'', which is equivalent to a normalized Fourier
transform~\cite{Moser}.
The oscillation amplitude $A$ is expected to be $A=0$ ($A = 1$)
for oscillation frequencies sufficiently far from (close to) the true
value of $\dms$.
All available measurements of the oscillation amplitude at $\dms = 15~\ips$ are
summarized in Figure~\ref{fig_ampW}. Also shown are the sensitivities for each
analysis to set a 95\% C.L. lower limit on $\dms$.

\begin{figure}
\begin{center}
  \epsfxsize10cm
  \epsfbox{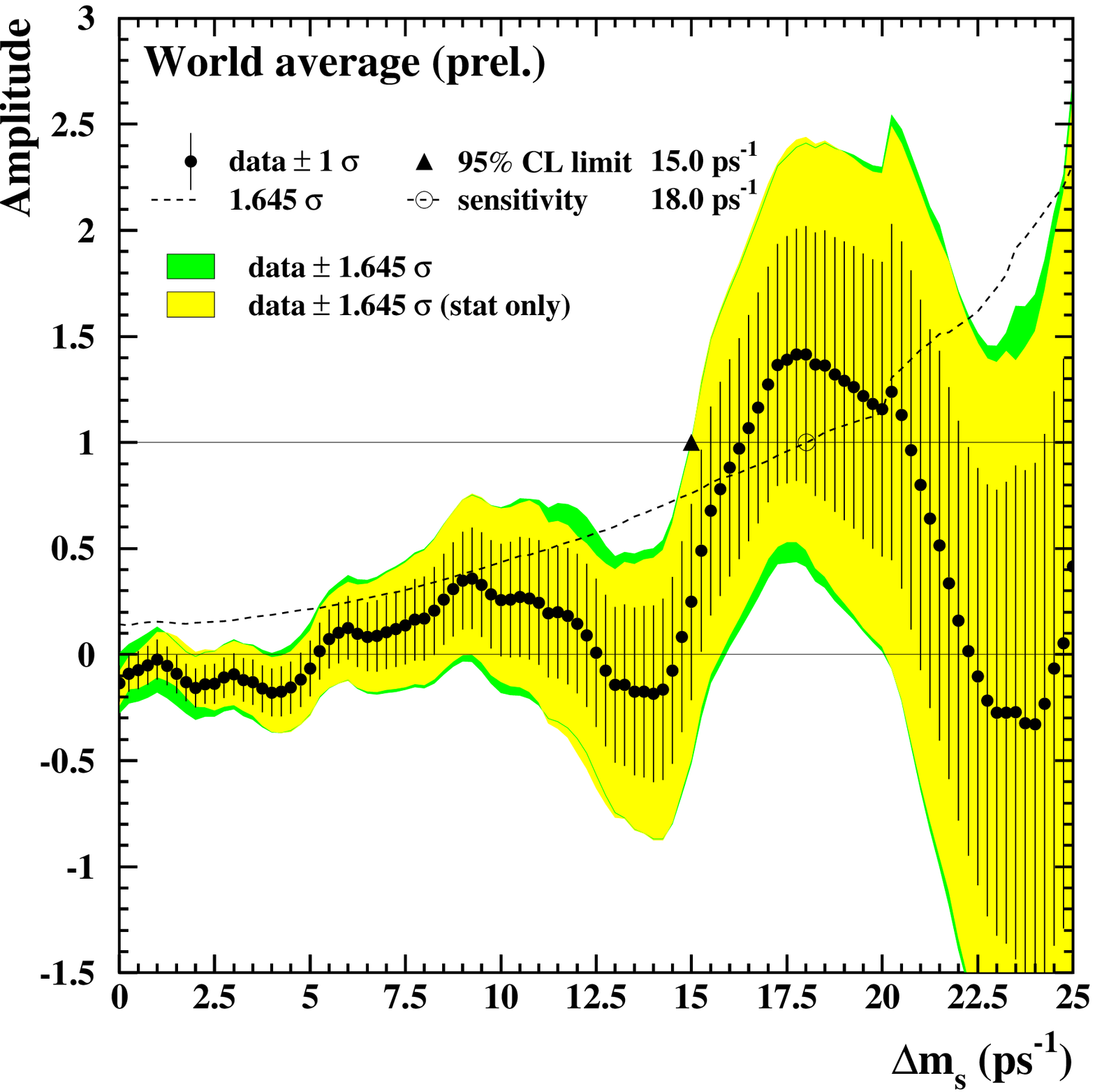}
\end{center}
\caption{World average $\bs$ oscillation amplitude as a function of
         $\dms$. }
\label{fig_afitW}
\end{figure}

  The measured oscillation amplitudes are combined \cite{BOSCWG},
taking statistical and systematic
correlations into account, to obtain the world average amplitude spectrum
shown in Figure~\ref{fig_afitW}.
The rise in statistical error as $\dms$ increases comes from the fact
that an increasingly smaller fraction of the data sample has sufficient
proper time resolution to resolve more rapid oscillations;
the better the resolution, the smaller the rise.
The combined amplitude spectrum excludes mixing $(A = 1)$ for
$\dms < 15.0~\ips$ at the 95\% C.L., whereas the sensitivity is 18.0 $\ips$.
The significance of the deviation from $A = 0$ near $\dms = 17.5~\ips$
was investigated with a parameterized fast Monte Carlo simulation. It was
found that, assuming that the true value of $\dms$ is large,
the probability to observe a deviation of equal or greater significance
as that seen in figure~\ref{fig_ampW} anywhere
between $\dms$ of 0 and 25 $\ips$ is about 2.5\%.

Further progress is still expected from DELPHI and SLD
over the next year or so.
In the near future, CDF (and presumably D0) is expected to apply
the fully exclusive method and observe $\bs$ oscillations with $5 \sigma$
significance up to $\dms$ of 40 $\ips$.
Beyond this, both BTeV and LHC-b are expected to measure $\dms$
with statistical precision of 0.1\% or better.

\section{Width difference in \bes\ mesons}

As discussed in the previous section, the $\bsmix$ system consists of
mass eigenstates $\rm B_{s}^{light}$ and 
$\rm B_{s}^{heavy}$. Neglecting CP violation,
these are also CP eigenstates \bshort(CP=$+1$) and \blong(CP=$-1$) with
different decay widths \gshort\ and \glong, since the quark level \bes\ decay
process $\rm b(\bar s)\rightarrow c\bar cs(\bar s)$ gives rise to mainly
CP-even final states. The width difference $\delgams=\gshort-\glong$ could
be as large as 20\,\%, and a non-zero value would mean that the 
\bes\ meson would have two distinct components with different lifetimes
$\tshort=1/\gshort$ and $\tlong=1/\glong$. The corresponding width difference
in the \bzero\ meson system is expected to be negligible.

Experimentally, a non-zero value of \delgams\ can be searched for in two ways:
measuring the lifetime of a CP eigenstate decay mode and comparing the
result with the inclusive \bes\ lifetime, or looking for deviations from a 
single exponential in inclusive ({\em e.g.\/} semileptonic) \bes\ decays.
The former has a linear sensitivity to \delgams, but is limited by the small
branching ratios to CP eigenstates, whilst the latter method has larger
statistics but less sensitivity as the dependence on \delgams\ is only
second order.

Measurements of the first type include a CDF analysis of 
$\bes\rightarrow\rm J/\psi\phi$ \cite{cdfjpsi} and an ALEPH analysis of 
$\rm\bes\rightarrow\phi\phi X$ resulting from the dominantly CP-even
decay $\rm\bes\rightarrow D_s^{(*)+}D_s^{(*)-}$ \cite{alephdsds}. 
DELPHI have also measured
the \bes\ lifetime in $\bes\rightarrow\rm D_s X$, assumed to contain
more CP-even than CP-odd decays \cite{delphibsds}.

The \bes\ lifetime measurements from semileptonic \bes\ decays 
(see Section~\ref{rhs:taubs}) also provide a constraint on \delgams\ as a
function of the average \bes\ width $\rm\Gamma_s=(\glong+\gshort)/2$, since
the semileptonic width $\Gamma_{\rm sl}$ is expected to be the same
for both \bshort\ and \blong. This leads to different semileptonic branching
ratios $\Gamma_{\rm sl}\tshort$ and $\Gamma_{\rm sl}\tlong$ for the two 
states, and hence an unequal mixture of them in the semileptonic decay sample.
The sensitivity is enhanced if the average \bes\ lifetime 
$\rm\taubs=1/\Gamma_s$ is assumed to be equal to the \bzero\ lifetime, which 
is expected to be true to better than 1\,\%. Finally, L3 have set a direct
limit on \delgams\ by studying the proper time distribution in a sample of
topologically selected neutral b hadron decays (see section~\ref{rhs:taubp})
and assuming values for the \bzero\ and \bbary\ lifetimes \cite{l3top}.

The various results from LEP and CDF have been combined to give an overall
result of $\delgams/\Gamma_{\rm s}=0.16^{+0.08}_{-0.09}$ or 
$\delgams/\Gamma_{\rm s}<0.31$ at 95\,\% CL \cite{lephfst}. If the
$\taubz=\taubs$ assumption is removed, the result changes to 
$\delgams/\Gamma_{\rm s}=0.24^{+0.16}_{-0.12}$ or 
$\delgams/\Gamma_{\rm s}<0.53$ at 95\,\% CL. These results are 
suggestive of a significant non-zero value of \delgams, consistent with
the high limit on $\dms$, but are not yet significant enough to draw any
definite conclusions.


\section{Measurements of \mvcb}

The magnitudes of the CKM matrix elements \vcb\ and \vub\ are fundamental 
parameters of the Standard Model which can only be determined experimentally,
and are of particular interest in the CKM model of CP violation.
Their values govern the transition rates of the quark level processes 
$\rm b\rightarrow c\ell\bar\nu$ and $\rm b\rightarrow u\ell\bar\nu$. 
However, the confinement of quarks within hadrons leads to significant
theoretical uncertainties in interpreting the corresponding hadronic decays.

The magnitude of \vcb\ is measured from the differential rate
of \bztodslv\ decays as a function of the recoil variable $\omega$, defined 
as the product of the four-velocities of the B and \dstar\ mesons. $\omega$
varies from 1 at the point of zero recoil (when the \dstar\ is produced
at rest in the B rest frame) to about 1.5. The differential decay rate is 
given by
\[
\frac{d\Gamma}{d\omega}=\frac{1}{\taubz}
\frac{d\bratio{\bzero}{\dstar\ell\bar\nu}}{d\omega}=
{{\cal F}^2(\omega)}|\bvcb|^2{\cal K(\omega)}
\]
where \fw\ is the hadronic form factor, \kw\ is a known phase space term
and \taubz\ is the \bzero\ lifetime \cite{vcbth1,isgw}.
In the heavy quark limit, \fw\ corresponds to the Isgur-Wise function
and $\fone=1$ \cite{isgw}. Dispersion relations allow the form of 
\fw\ to be calculated
in terms of a single parameter $\rho^2$ which represents the slope of \fw\ at 
$\omega=1$ \cite{clnff}. The value of \fone\ can be calculated using
Heavy Quark Efective Theory (HQET);
one recent estimate gives $\fone=0.913\pm 0.042$ \cite{babar}. 
Unfortunately, the phase
space term $\kw\rightarrow 0$ as $\omega\rightarrow 1$, so the
differential decay rate must be measured close to $\omega=1$ and
extrapolated to determine \mvcb.

The signal decay can be reconstructed exclusively, where the \dzero\
is reconstructed in specific decay modes, or inclusively, as discussed
in section~\ref{rhs:taubp}.
The recoil $\omega$ is reconstructed for each event from the estimated
\dstar\ and \bzero\ four-momenta. The unmeasured neutrino is reconstructed
from the missing momentum vector in the event, and the \bzero\ 
flight direction  can also be inferred from its reconstructed secondary 
vertex position, particularly for long-lived \bzero\ decaying far from the
primary vertex. The resolution on the reconstructed $\omega$ is typically
0.07--0.15, and the reconstruction efficiency is almost flat as a function of 
$\omega$, allowing a precise extrapolation to $\omega=1$.

The values of \fvcb\ and \rhsq\ are extracted using likelihood fits
to the reconstructed $\omega$ spectra, taking into account contributions
from signal decays, physics background (mainly from
decays of the form \btodshlv) and combinatorial background. 
The combinatorial background is 
estimated using events from the mass sidebands or having the wrong
lepton-pion charge correlation. The physics background has contributions
from \btodsslv\ decays where the \ddstar decays to $\rm D^*\pi$ or $\rm D^*K$,
and possibly also from direct four-body non-resonant \btodshlv\ decays.
Isospin considerations imply that approximately two thirds of these decays
involve charged pions, which can be suppressed by looking for additional
charged particles consistent with coming from the b decay vertex.

The contribution of \btodshlv\ events can be calculated from the ALEPH
measurement of
$\bratio{\rm b}{\rm\dstar\pi^-\ell\bar\nu}=(0.473\pm 0.095)\,\%$ 
\cite{alephdss}, but the $\omega$ spectrum of these events is also important.
The decay rates and form factors
predicted by the calculation of Leibovich {\em et al.\/}\cite{ligeti} have
been used, with the free parameters of the calculation constrained by
experimental measurements of the ratio 
$R^{**}=\rm\Gamma(\bar B\rightarrow D_2^*\ell\bar\nu)/\Gamma(\bar B\rightarrow
D_1\ell\bar\nu)=0.37\pm 0.16$ \cite{lephfst}. 

A recent preliminary DELPHI analysis \cite{delphidss}
has searched for mass structure in 
$\rm\bar B\rightarrow\dstar\pi^-\ell\bar\nu$, 
$\rm\bar B\rightarrow\dzero\pi^+\ell\bar\nu$ and
$\rm\bar B\rightarrow\rm D^+\pi^-\ell\bar\nu$ 
decays, by combining charged pions with exclusively reconstructed
$\rm D^{(*)}\ell$ combinations. DELPHI  measure branching ratios for 
$\rm\bar B\rightarrow D_1\ell\bar\nu$ and 
$\rm\bar B\rightarrow D_2^*\ell\bar\nu$ consistent with earlier results
from ALEPH and CLEO, and also reports first evidence for the production
of the broad
$\rm D_1^*$ state, $\bratio{\rm\bar B}{D_1^*\ell\bar\nu}=(1.63\pm 0.55)\,\%$,
in agreement with expectations from HQET. Since the understanding of the
background from these decays is the largest systematic error on the LEP 
measurements of \mvcb, further results in this area would be extremely
beneficial.

The fit results for \fvcb\ and \rhsq\ from the ALEPH exclusive \cite{alephvcb},
DELPHI inclusive \cite{delphivcb} and OPAL inclusive and exclusive combined 
\cite{opalvcb} analyses are shown 
in Figure~\ref{rhf:vcbres}. Some of these results used linear or quadratic
parameterisations of the hadronic form factor \fw, and older evaluations
of the background from \btodshlv\ decays.
The central values and uncertainties of all the
measurements have been corrected by the LEP \vcb\ working group to use the
latest form factor and background evaluations. The combined LEP result, taking
into account correlated systematic uncertainties, is
\begin{eqnarray*}
\fvcb & = & (34.9\pm 0.7\pm 1.6)\times 10^{-3} \\
\rhsq & = & 1.12\pm 0.08\pm 0.15 
\end{eqnarray*}
where the first error is statistical and the second systematic in each case
\cite{lephfst}.

\rhepostfig{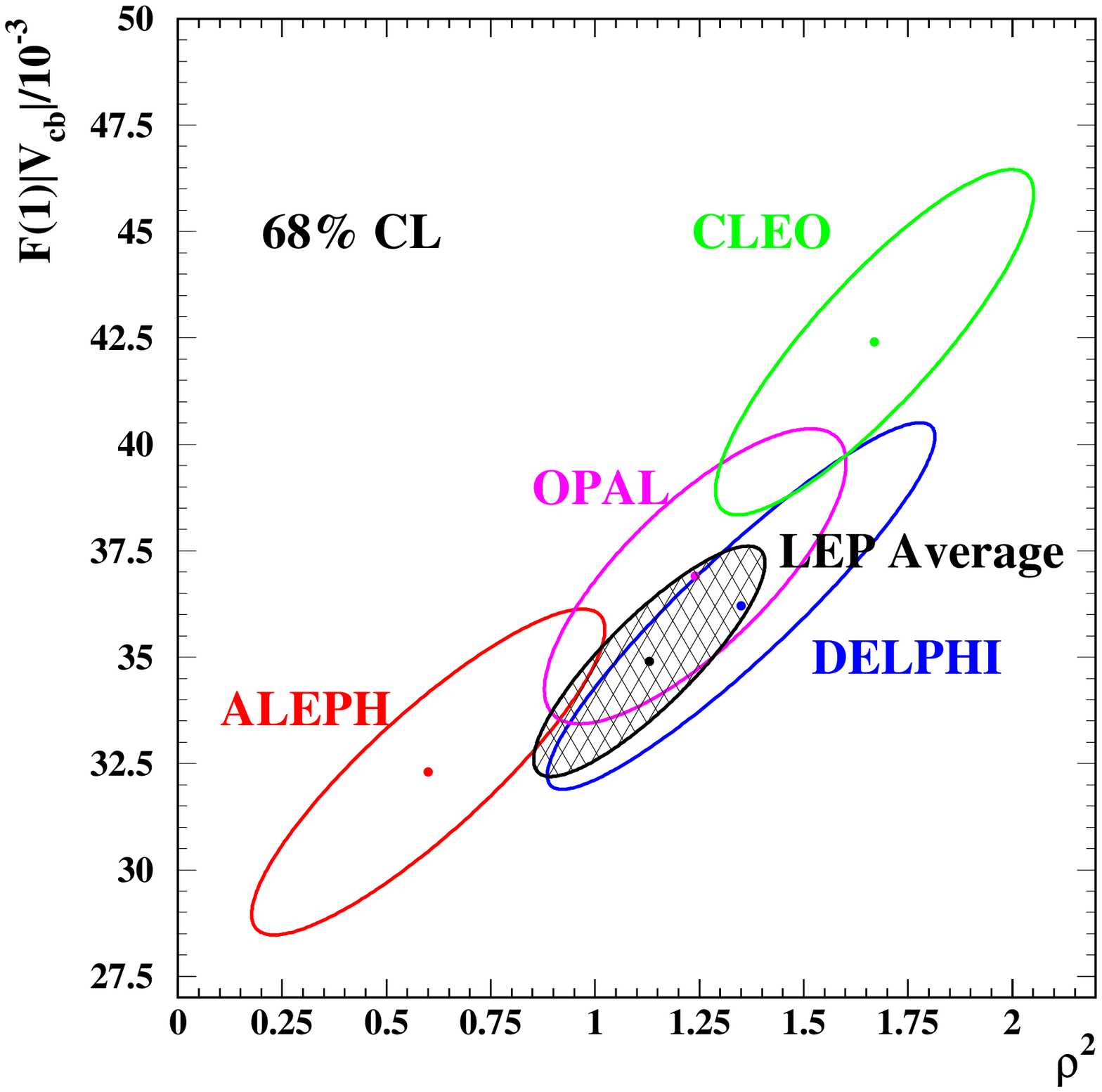}{rhf:vcbres}{0.2}{0.6}
{Experimental measurements of \fvcb\ and \rhsq\ from 
ALEPH, DELPHI, OPAL and CLEO.}

The recent measurement from CLEO \cite{cleovcb} is also shown in 
Figure~\ref{rhf:vcbres}, and is somewhat
higher than the LEP result. Assuming no correlated systematic errors, and 
correcting to a common form factor parameterisation, the LEP and CLEO
\fvcb\ results differ by 2.4 standard deviations. More work is needed to 
understand the origin of this possible discrepancy.

Using the value $\fone=0.913\pm 0.042$ \cite{babar}, the LEP combined result is
\[
\mvcb = (38.2\pm 1.9\pm 1.8)\times 10^{-3}
\]
where the first error is the experimental error on \fvcb\ and the second
the theoretical uncertainty on the value of \fone.

The value of \mvcb\ can also be extracted from the LEP measurement of the
inclusive semileptonic branching ratio 
$\bratio{\rm b}{\rm X\ell\bar\nu}=(10.56\pm 0.11\pm 0.18)\%$ \cite{lepew},
after subtracting the contribution from $\rm b\rightarrow u\ell\bar\nu$.
Theoretical models can then be used to translate this into a measurement of
\mvcb\ \cite{lephfst}, giving 
$\mvcb\rm =(40.7\pm 0.5\,(expt) \pm 2.0\,(theory))\times 10^{-3}$,
in encouraging agreement with the determination from \bztodslv\ decays.
Significant correlations (both experimental and theoretical) exist between
these measurements, so no attempt is made to average them at present.

\section{Measurements of \mvub}

The magnitude of \vub\ can be measured from the rate of charmless 
semileptonic b decays $\rm\bar B\rightarrow X_u\ell\bar\nu$, but there is an 
overwhelming background from charmed semileptonic b decays, due to the
smallness of the ratio $(\mvub/\mvcb)^2\sim 10^{-2}$. The main challenges of
this measurement are therefore to separate a clean sample of 
charmless decays whilst maintaining a model independent efficiency and good 
control over the charmed background.

Previous measurements of \mvub\ at the $\Upsilon(4S)$ resonance have used 
either a lepton endpoint technique (selecting events with lepton energies
beyond the kinematic limit for $\rm\bar B\rightarrow D\ell\bar\nu$ decays) 
\cite{endpvub}, or
the reconstruction of specific exclusive final states 
($\rm\bar B\rightarrow \pi\ell\bar\nu$ or $\rho\ell\bar\nu$) \cite{cleovub}.
The former 
suffers from significant dependence on the modelling of the lepton energy
spectrum, whilst the latter requires an estimate of the decay form factor 
in a regime where HQET is not applicable.

The experimental environment at LEP allows a rather different and complementary
technique to be used, based on inclusive reconstruction of the 
hadronic system recoiling against the lepton. In most 
$\rm\bar B\rightarrow X_u\ell\bar\nu$ decays, the invariant mass of this
system is significantly below the charm hadron mass, allowing a relatively
model independent measurement of the charmless semileptonic branching ratio.
Theoretical models can then be used to relate this branching ratio to
$\mvub^2$.

The ALEPH \cite{alephvub} and L3 \cite{l3vub}
analyses start by selecting a sample of events with an identified
high momentum lepton (electron or muon) emitted with significant transverse
momentum with respect to the nearest jet axis. Secondary vertex based b-tagging
is used to suppress non-\bbbar\ event background. Reconstructed particles 
(tracks and calorimeter clusters) in the jet containing the lepton are then
classified as coming from the b hadron decay or fragmentation processes,
based on momentum, rapidity and vertexing information. The b decay particles
are boosted into the reconstructed b hadron rest frame, and various
kinematic and event shape variables, as well as the invariant
mass of the hadronic system, are calculated.
Both charged and neutral particles are used,
to minimise the dependence on the particular b decay mode. The final selection
is typically made using an artificial  neural network, trained to separate
$\rm\bar B\rightarrow X_u\ell\bar\nu$ decays from 
$\rm\bar B\rightarrow X_c\ell\bar\nu$
decays using all available discrimination variables and their correlations.

A somewhat different approach is taken by DELPHI \cite{delphivub}. Here, the
selected events are divided into 4 classes, depending on the
reconstructed hadronic mass (above or below 1.6\,GeV) and an enrichment
or depletion in $\rm b\rightarrow u$ decays. The enrichment is performed
by selecting events with no identified kaons or protons (from charm decays), 
and by trying to
associate the lepton with the secondary vertex of the hadronic system---in
$\rm\bar B\rightarrow X_c\ell\bar\nu$ decays, the lepton tends to be produced 
closer to the primary vertex due to the flight distance of the charm hadron 
before it decays. An excess of events is found in the class with
low hadronic mass and $b\rightarrow u$ enrichment, consistent with the
presence of $\rm\bar B\rightarrow X_u\ell\bar\nu$ decays. 
The analysis measures
the ratio $\bratio{\bar B}{X_u\ell\bar\nu}/\bratio{\bar B}{X_c\ell\bar\nu}$
using a binned maximum likelihood fit to the number of events and lepton
energy spectrum of each class.

The measurements from ALEPH \cite{alephvub}, DELPHI \cite{delphivub} and 
L3 \cite{l3vub} have been combined to give a result of
\[
\bratio{\rm\bar B}{\rm X_u\ell\bar\nu}=(1.74\pm 0.37 {\rm (expt)}\pm 0.38
 {\rm (b\rightarrow c)}\pm 0.21 {\rm (b\rightarrow u)} )\times 
10^{-3}
\]
where the first error includes statistical and uncorrelated systematic 
contributions, the second is associated with modelling $\rm b\rightarrow c$
decays and the third with modelling $\rm b\rightarrow u$ decays \cite{lephfst}.

This result can be translated into a value for \mvub\ using the relation
\cite{vubtheory}
\begin{eqnarray*}
\mvub & = & 0.00445\times\left(\frac{\rm BR(\btoxulv)}{0.002}
\frac{1.55\,\rm ps}{\tau_{\rm b}}\right)^{\frac{1}{2}}\\
 & & \times
{(1\pm 0.02{\rm (pert)}\pm 0.035(\rm m_{\rm b}))}
\end{eqnarray*}
where $\tau_{\rm b}$ is the inclusive b hadron lifetime and the uncertainties 
refer to unknown higher order corrections and the value of the b quark mass.
Using $\tau_{\rm b}=1.564\pm 0.014$\,ps \cite{lephfst} this gives
\[
\mvub=\left(4.13^{+0.42}_{-0.47}\,{\rm (expt)}\,{^{+0.43}_{-0.48}
{\rm (b\rightarrow c)}}{^{+0.24}_{-0.25}{\rm (b\rightarrow u)}}
{\pm 0.20{\rm (theory)}}\right)\times 10^{-3} \ ,
\]
where the errors correspond to  uncorrelated experimental, 
$\rm b\rightarrow c$, $\rm b\rightarrow u$ and theoretical sources 
respectively. As expected, the largest error comes from the uncertainties
in modelling the large background from $\rm\bar B\rightarrow X_c\ell\bar\nu$
decays in these analyses.

\section{Conclusions}

The LEP and SLD experiments have made an enourmous contribution to the 
knowledge of b physics. The individual b hadron lifetimes have each been
measured to 1--4\,\%, the \bzero\ oscillation frequency has been measured
to 3\,\% and an impressive lower limit of 15.0\,$\rm ps^{-1}$ has been 
set on the \bes\ oscillation frequency. Limits have also been set on
the  $\bes-\bsbar$ width difference $\delgams$. Finally, the CKM element
\mvcb\ has been measured to 7\,\%, and evidence has been seen for a non-zero
value of \mvub. With the exception of the unexpectedly low value of the
\bbary\ lifetime, all measurements are in impressive and consistent agreement
with theoretical expectations and the Standard Model CKM mechanism of 
CP violation.


\end{document}